\documentclass{aip-cp}

\usepackage[numbers]{natbib}
\usepackage{rotating}
\usepackage{graphicx}
\usepackage{xcolor}
\begin{document}

\title{The Cosmological Lithium Problem Revisited}

\author[aff1,aff2]{C.A. Bertulani\corref{cor1}}
\author[aff2]{A.M. Mukhamedzhanov}
\eaddress{akram@comp.tamu.edu}
\author[aff1]{Shubhchintak}
\eaddress{shub.shubhchintak@tamuc.edu}
\affil[aff1]{Department of Physics and Astronomy, Texas A\&M University-Commerce, Commerce, TX 75429, USA}
\affil[aff2]{Department of Physics and Astronomy, Texas A\&M University, College Station, TX 75429, USA}
\affil[aff3]{Departamento de F\'\i sica, Universidade de Coimbra, 3004-516 Coimbra, Portugal}
\corresp[cor1]{carlos.bertulani@tamuc.edu}

\maketitle

\begin{abstract}
After a brief review of the cosmological lithium problem, we report a few recent attempts to find theoretical solutions by our group at Texas A\&M University (Commerce \& College Station). We will discuss our studies  on the theoretical description of  electron screening, the possible existence of parallel universes of dark matter, and  the use of non-extensive statistics during the Big Bang nucleosynthesis epoch. Last but not least, we discuss possible solutions within nuclear physics realm. The impact of recent measurements of relevant nuclear reaction cross sections for the Big Bang nucleosynthesis based on indirect methods is also assessed. Although our attempts may not able to explain the observed discrepancies between theory and observations, they  suggest theoretical developments that can be useful also for stellar nucleosynthesis.
\end{abstract}

\section{INTRODUCTION}

The so-called lithium problem is one of the most resilient puzzles in nuclear astrophysics. It is quickly becoming as well known as the solar neutrino problem which ruled for many years as a seed for the development of theoretical physics imagination \cite{BP92,BB93,Hax95,Bah97}. It eventually was solved with an old idea: that neutrinos can oscillate \cite{Pon57}. The solar neutrino problem is arguably one of the main reasons for the strong investment on neutrino astrophysics during the last decades, which ultimately led to two neutrino physics related nobel prizes a few years back. It was a sad decision by the Nobel committee that  John Bahcall was not awarded with this prize as he was the most creative physicist in neutrino astrophysics before and after the very first design of the Homestake experiment \cite{Hax09}. Among several possibilities to solve the puzzle was thought to be in nuclear physics, e.g.,  the $^7$Be(p,$\gamma$)$^8$B reaction, responsible for the high energy neutrinos coming from the core of the sun \cite{Bert95,Tau94}. But only a drastic modification of the low energy astrophysical S-factor for this and other reactions would solve the solar neutrino puzzle. The solution came indeed from the unusual character of the neutrino as a fundamental particle, which took a long time to be proven experimentally.

In contrast to the solar neutrino problem, there are no hints that the cosmological lithium problem can be solved without closely assessing the role of  nuclear reactions, during the Big Bang, or in stellar environments. A few minutes ($\sim 4 - 20$ m) after the Big Bang, nuclei started forming first with the creation of the deuteron by neutron capture on proton, p(n,$\gamma$)d. The formation of deuterons is strongly dependent on the   baryon-to-photon ratio in the Big Bang epoch,  $\eta_{b}=\#baryons/\#photons$.  After the deuteron bottleneck is surpassed,   all other heavier elements are synthesized, and are also therefore strongly dependent on $\eta_{b}$.  Other important parameters of Big Bang nucleosynthesis are the neutron lifetime and the number of neutrino species. Because deuterons are promptly consumed once they are formed, a significant amount of $^3$He is created  by means of the (p,$\gamma)^3$He and d(d,n)$^3$He reactions. Tritium is also synthesized by means of d(d,p)t reaction. The creation of $^4$He nuclei then follows  by the $^3$He(d,p)$^4$He and t(d,n)$^4$He reactions. After hydrogen (75\%), helium (25\%) is the most abundant element in the visible universe. The correct prediction of the hydrogen and helium abundance is one of the major successes of the Big Bang model. It is necessary to have a neutron-to-proton ratio, n/p = 1/7, when the Big Bang nucleosynthesis started in order to explain the observed values of hydrogen and helium abundance. Thus, Big Bang nucleosynthesis occurred in a proton-rich environment. 

A series of reactions involving neutron, proton, deuteron, and helium captures allow elements up to lithium and beryllium to be created during the Big Bang.  ${}^{7}{\rm Li}$ and  ${}^{6}{\rm Li}$, were synthesized in very small amounts, with the $^7$Li/H abundance of the order of $10^{-10}$ and $^6$Li/H of the order of $10^{-14}$.  Not only the deuteron, but also the $^7$Li abundance is strongly dependent on the value of $\eta_b$. Big Bang nucleosynthesis is not the only source of lithium isotopes. $^6$Li can also be produced in cosmic rays due to spallation reactions and $^7$Li can be formed in novae and pulsations of AGB stars. Spite and Spite \cite{Spi82a,Spi82b} have noticed that the $^7$Li abundance was nearly independent of the metallicity in metal-poor stars. Metal poor stars are defined as those with a small Fe/H abundance relative to the sun. The logarithm of this ratio is denoted by [Fe/H]. In low metallicity stars $-2.4\leq [{\rm Fe/H}] \leq-1.4$. The observations were done for warm ($5700 \leq T  \leq 6250$ K)  metal-poor dwarf stars. The nearly constant $^7$Li abundance, nearly independent of metallicity and temperature, is known as the Spite plateau. White dwarfs at moderate temperatures have been used in such studies because warmer stars, such as red giants, reaching temperatures in excess of $10^6$ K lead to the destruction of $^7$Li via the reaction ${}^{7}{\rm Li}({\rm p},\alpha){}^{4}{\rm He}$.  Based on the Spite plateau observations, a reasonable conclusion is that lithium is not created or depleted in warm dwarfs, even over a rather large variation of temperature. It is thus natural to conclude that the Spite plateau consists in a firm observation of the primordial $^7$Li abundance, synthesized during the Big Bang epoch.  But it is important to notice that there have been recent observations in low-metallicity stars which seem to be at odds  with the Spite plateau \cite{Aok09,Sbo10}. 

Big Bang nucleosynthesis leads to robust predictions which have survived the test of observations.  Observations of lithium abundance in metal poor halo stars \cite{Sbo10}  yield $^{7}{\rm Li/H} = 1.58^{+0.35}_{ - 0.28} \times  10^{-10}$ which is appreciably smaller than the value of ${}^{7}{\rm Li/H}=4.46 \times 10^{-10}$ predicted by Big Bang Nucleosynthesis (BBN) \cite{Piz14}. This difference has survived a large number of tests and is the source of the lithium puzzle.  An additional problem is the abundance of ${}^{6}{\rm Li}$  created in the BBN by means of the $^{2}{\rm H}(\alpha,\gamma)^{6}{\rm Li}$ reaction. When $^{6}{\rm Li}$ is formed within stars, it disappears quickly by means of other nuclear reactions. One believes that it is mainly created  by  cosmic rays, but one also believes that it can exist in the atmosphere of  halo metal-poor warm dwarfs, surviving destruction by cosmic rays. But this assumption is controversial, as the same hypotheses can be applied to ${}^{7}{\rm Li}$ nuclei. While the  BBN predicts a primordial isotopic ratio  ${}^{6}{\rm Li}/{}^{7}{\rm Li}  \sim  10^{-5}$ \cite{Piz14}, some observations report a value $^{6}{\rm Li}/{}^{7}{\rm Li}  \sim 5 \times 10^{-2}$ \cite{Asp06}. This discrepancy is an addition to the BBN lithium puzzle and is known as the second lithium problem.  Such findings have been disputed and the complexity of 3D models for convection and non-local thermodynamical equilibrium in the photo-sphere of metal-poor stars can lead to a substantial weakening of the second lithium problem, leading to isotopic ratios in better agreement with the BBN predictions \cite{LMA13}.  

\section{ENVIRONMENT ELECTRONS}\label{enve}

It is well known that nuclear reaction cross sections at the low energies of interest for astrophysics are enhanced because of the presence of electrons in the environment. Electrons bound in the atoms as well as free electrons, as those present in stellar plasmas, reduce the Coulomb repulsion between nuclei and enhance the chances for tunneling through the Coulomb barrier. There have been experimental evidences that the effect of atomic electrons is not well described by theory  \cite{BG10}. In laboratory experiments, the cross section enhancement due to bound electrons in targets is larger than predicted by apparently all existing models in theory \cite{BBH97}.  In stars, or during the Big Bang, the cross section enhancement due to free electrons is quantified by the enhancement factor $  f(E) = {\sigma _{s}(E)/ \sigma _{b}(E)}$, where $\sigma _{s}$ is the screened cross section and $\sigma _{b}$ is the bare cross section, without influence of electrons. This factor is usually calculated using the  Debye-H\"uckel approximation, yielding a screened Coulomb potential in the weak screening limit (for $\langle V\rangle \ll kT$),  so that $V(r) = ({e^2 Z_i / r}) \exp\left(-{r/ R_D} \right)$. This leads to an enhancement factor \cite{BG10}
\begin{equation}
    f =\;\mathrm{exp} \left({Z_1 \,Z_2 \;e^2 \over R_D k T}\right)
    =\;\mathrm{exp} \left( {0.188\,Z_1 \,Z_2 \;\zeta \;\rho
    ^{1/2}\;T_6^{-3/2} } \right),
        \label{f02}
\end{equation}
where the Debye radius is given by
\begin{equation}
R_D = {1\over \zeta}\left( {k T \over 4 \pi e^2 n} \right)^{1/2},
\label{DR}
\end{equation}
with the number density  given by $n$, the density $\rho$ given in units of g/cm$^3$, and
\begin{equation}\zeta = \left[ \sum\limits_i~X_i ~
\frac{Z_i^2}{A_i} +  \chi \sum\limits_i~X_i \frac{Z_i}{A_i} \right]^{1/2}.\end{equation}
In this equation, the mass fraction of particle $i$ is given by $X_i$, the temperature $T_6$ is given  in units of 10$^6$ K, and
$\chi$ is a factor correcting for  electron degeneracy effects \cite{Sal54}. 

There have been some discussions in the literature about the magnitude of screening effects and deviations form the Debye-H\"uckel model
\cite{Sha96,Sha00,Mao09,WFT01}. Applications of this approximation to the plasma in the core of the sun yields 3-5 of particles within the Debye-H\"uckel. This does not validate the use of mean-field approximations inherent to the derivation of the Eq. (\ref{f02}). A better method to describe the screening in such cases is the use of molecular dynamics simulations \cite{Sha96,Sha00} but these do not seem to reproduce the mean field models.  Electron screening is also dependent on the electron velocities in the plasma  \cite{KSK05}. Some lingering questions remain on the use of the Salpeter formula, Eq. (\ref{f02}), used to explain the experiments on electron screening \cite{RS95,Rol01}.  Other theoretical methods have also defied the traditional dependence on the Debye-H\"uckel approximation to describe electron screening in stars \cite{Lio03,Mic15}. Based on this discussion and the theoretical difficulty to reproduce the experimental data on laboratory screening of the astrophysical nuclear cross section and explain the stellar plasma screening at content, it is not far-fetched to assume that there might be strong deviations of the Debye-H\"uckel approximation. In Ref. \cite{Wan11} this possibility was explored in order to verify if electron screening could modify appreciably the abundance of elements formed in the BBN (see also, Ref. \cite{Itoh97}). 
 
\begin{center}
\begin{figure}[t]
{\includegraphics[width=8cm]{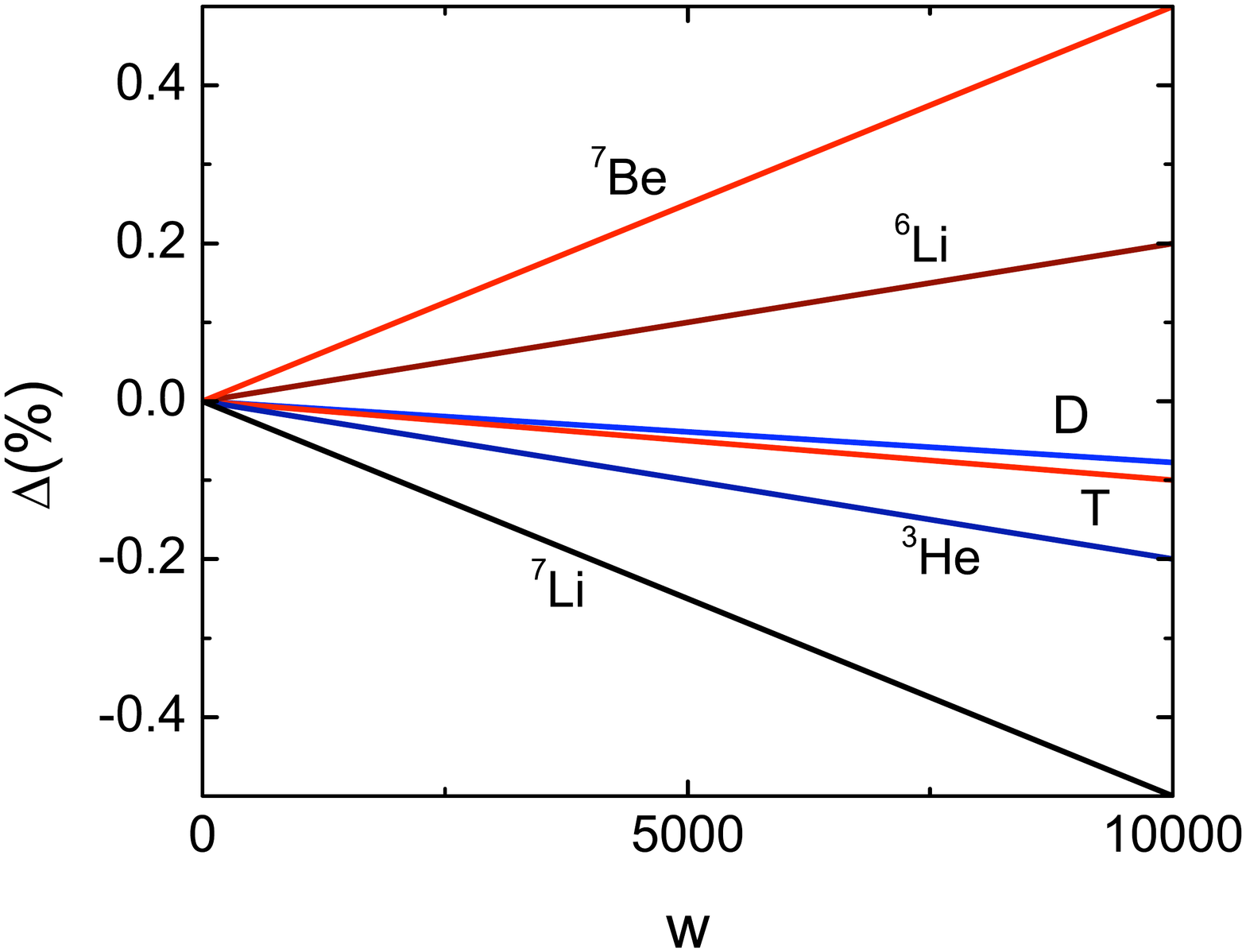}}
\includegraphics[width=9cm]{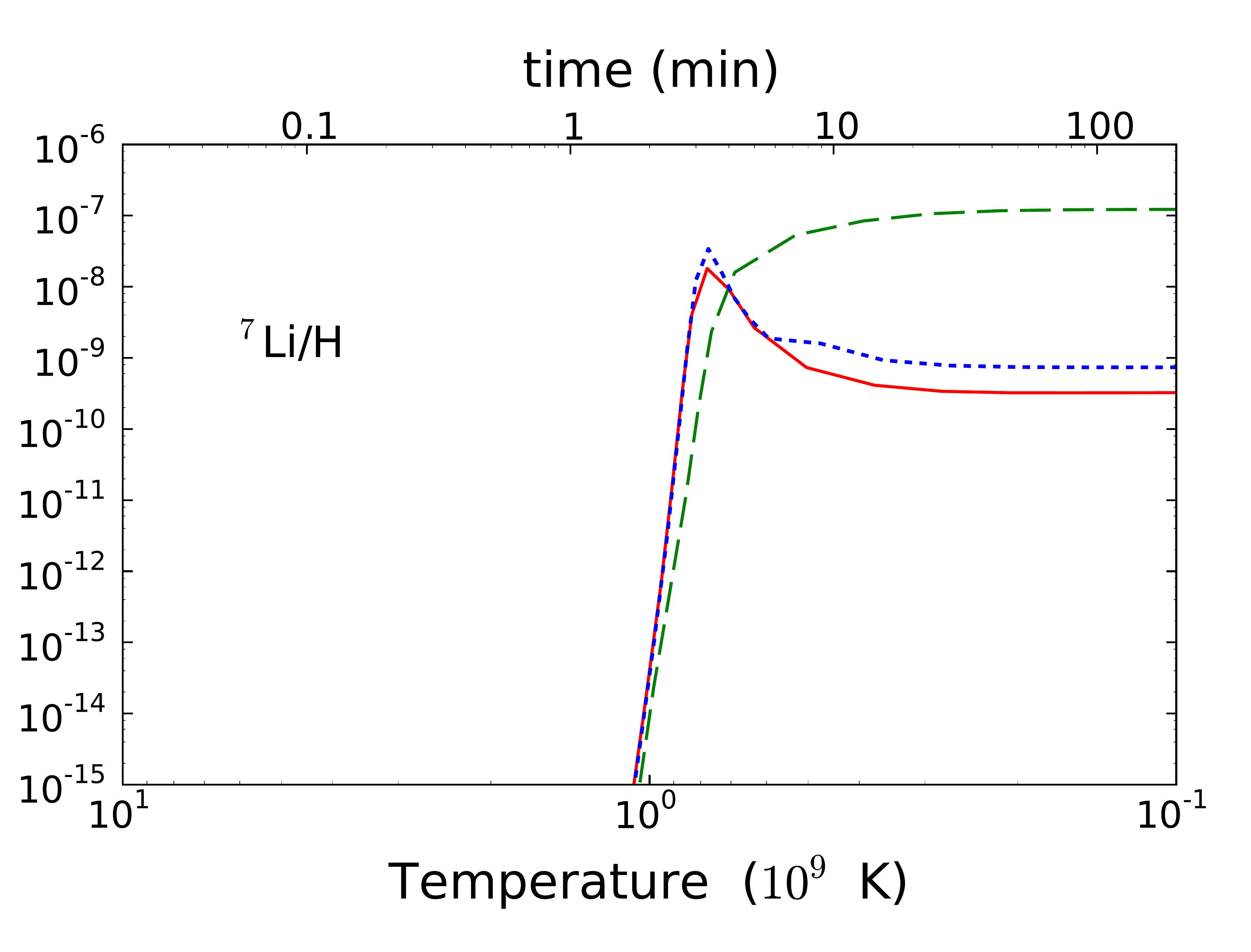}
\caption{
{\it Left:} Percent change in the abundances of  light nuclei synthesized during the BBN as  a function the fudge factor $w$ modifying the Debye-H\"uckel screening formula (Adopted from Ref. \cite{Wan11}).  {\it Right:} $^7$Li abundance in the BBN using reaction rates calculated with Maxwell distributions (solid curves) and with non-extensive statistics with $q=0.5$ (dotted line) and $q=2$ (dashed line)  (Adopted from Ref. \cite{BFH13}).  
}\label{fig1}
\end{figure}
\end{center}

In the early universe, the electron density decreased strongly as the temperature decreased and it was much larger (by up to $10^4$ times) than the density in the core of the sun ($n_e^{sun}\sim 10^{26}$/cm$^3$). The  baryonic density depends on the temperature as  $\rho_b \simeq h T_9^3$ ($T_9$ is the temperature in units of $10^9$ K), with $h$ being the baryon density parameter \cite{SW77}. $h$ changes value around $T_9\sim 2$ from $h\sim 2.1\times 10^{-5}$ to $h\sim 5.7\times 10^{-5}$. During the BBN the electron density was much larger than in the sun, but the baryon density was much smaller. The number of excess electrons is nearly the same as those of protons, with most of the remaining ones being equal to the number of positrons produced via  $\gamma \gamma \rightarrow e^+e^-$ processes.  From these considerations, one can calculate the enhancement factor in Eq. (\ref{f02})  and show that it becomes independent of the temperature \cite{Wan11}, 
\begin{equation}
f^{BBN}=\exp \left( 4.49 \times 10^{-8} \zeta Z_1 Z_2 \right)
\sim 1+4.49 \times 10^{-8} \zeta Z_1 Z_2  \ \ \ {\rm for} \ \ T_9\lesssim  1,
\label{fbbn}
\end{equation}
and
\begin{equation}
f^{BBN}=\exp \left( 2.71 \times 10^{-8} \zeta Z_1 Z_2 \right)
\sim 1+2.71 \times 10^{-8} \zeta Z_1 Z_2  \ \ \ {\rm for} \ \ T_9\gtrsim  2.
\end{equation}
In both situations the screening effect is vey small, unless one appreciably modifies the Deby-H\"uckel approximation. 

In Ref. \cite{Wan11} the Debye-H\"uckel approximation was modified  by including  an artificial  factor $w$, so that
\begin{equation}
f'=\;\mathrm{exp} \left(w{Z_1 \,Z_2 \;e^2 \over R_D k T}\right), \ \ \ \ \  {\rm or}  \ \ \ \ln f' =w\ln f,\label{w}
\end{equation}
where $1\leq w\leq 10^4$. The modified BBN abundances, $Y'$ were calculated using a standard BBN model \cite{Wan11} and the relative change $ \Delta = {(Y' -Y)/ Y}$ is shown in Figure \ref{fig1}.  It is seen that as $w$ increases, the BBN  abundances for $^6$Li and $^7$Be increase, but the abundances for D, T, $^3$He and $^7$Li decrease. But even if $w$ is as large as $10^4$ the modification of the BBN abundances is minimal. Thus, one concludes that the electron screening does not change BBN abundances appreciably and much less it can be identified as a solution of the lithium problem.

\section{NON-EXTENSIVE STATISTICS}

For the calculation of reaction rates in the BBN  one assumes the validity of the Maxwell-Boltzmann (MB) distribution of velocities of the nuclei in a plasma.  The MB distribution is based on the Boltzmann-Gibbs statistics, which assumes that (a) the average time between collisions is much larger than the collision time, (b) the interaction occurs locally, (c) there is no correlation between the velocities of two particles in a given location in space, and (d) the energy is conserved without transfer of energy from collective degrees of freedom. These are strongly constraining assumptions which do not always hold for systems in thermodynamical equilibrium.   Variations of the Boltzmann-Gibbs (BG) statistics have been proposed in Refs. \cite{Ren60,Ts88,GT04}. In Ref. \cite{BFH13}, the departure from the extensive BG statistics has been studied in the context of the BBN. The motivation was again to check if this would have any visible effect on the lithium abundances and a possible relation to the lithium puzzle. We  adopted the non-extensive statistics proposed by Tsallis \cite{Ts88,GT04}, which encompasses a large family of entropies, each depending on a parameter $q$, which serves as a measure of the departure from the extensive BG statistics. When  $q = 1$ the BG statistics is recovered from the Tsallis statistics. 

\begin{center}
\begin{figure}[t]
 
\end{figure}
\end{center}

The standard BBN model has been very successful and one cannot expect that a large departure from the BG statistics is possible. That is, the non-extensive parameter $q$ cannot be much different than $q=1$. In Ref. \cite{BFH13}, the value of $q$ was varied substantially so that the departure from the BBN predicted abundances can become visible. The deviation from the Maxwellian velocity distribution and its consequences for nuclear burning in stars has also been investigated in Refs.  \cite{MQ05,HK08,Deg98,Cor99}.  The entropy in the BG statistics is given by $ {\cal S}_{BG}=-k_B \sum_i p_i \ln p_i$,  where $p_i$ is the probability to find the system in the i-th microstate. If $A$ and $B$  are two independent systems,  the probability of  $A+B$ to be found in a state $i+j$,  is $ p^{A+B}_{i+j} = p^A_i \cdot p^B _j$. This yields the extensive property of the BG entropy,   ${\cal S}_{A+B}={\cal S}_A +{\cal S}_B$. In this case, a non-Maxwellian velocity distribution will be valid for particles in a plasma. For a Maxwellian distribution, the reaction rate in stellar plasmas is assumed to be
\begin{equation}
r_{ij}=\frac{N_{i}N_{j}}{1+\delta_{ij}} \langle \sigma v \rangle = \frac{N_{i}N_{j}}{1+\delta_{ij}}\left(\frac{8}{\pi\mu}\right)^{\frac{1}{2}}\left(\frac{1}{k_BT}\right)^{\frac{3}{2}}
\int_{0}^{\infty}dES(E) \exp\left[-\left(\frac{E}{k_BT}+2\pi\eta(E)\right)\right], \label{rij}
\end{equation}
with $\sigma$ being the cross section, $v$ the relative velocity of the particles $i$, $j$, $N_i$ the number of particles $i$, $\mu$  the reduced mass, $T$ the temperature, and $S(E)$ the astrophysical S-factor. $\eta=Z_iZ_je^2/\hbar v$ is the Sommerfeld parameter, for charges $Z_i$ and $Z_j$ and the relative energy of $i+j$ is given by $E=\mu v^2/2$. The astrophysical S-factor is defined as $ S(E)=E\sigma(E) \exp\left[{2\pi\eta(E)}\right]$. For neutron induced reactions it is more appropriate to use the definition $\sigma(E)= {R(E) / \sqrt{E}}$ because the  neutron-nucleus cross sections at low energies typically behave as $\sigma \propto 1/v$.

The effects of a non-extensive statistics for  BBN has been used to make predictions for the abundances of  ${^4}$He, D, ${^3}$He, and ${^7}$Li  and for the reaction rates of p(n,$\gamma$)d, d(p,$\gamma){^3}$He, d(d,n)${^3}$He, d(d,p)t, ${^3}$He(n,p)t, t(d,n)${^4}$He, ${^3}$He(d,p)${^4}$He, ${^3}$He$(\alpha,\gamma){^7}$Be, t$(\alpha,\gamma){^7}$Li, ${^7}$Be(n,p)${^7}$Li and ${^7}$Li(p,$\alpha){^4}$He using the available experimental data of these reactions \cite{BFH13}.  As an example, Figure \ref{fig1} (right) shows the $^7$Li abundance, where the solid curve uses Maxwell distributions to calculate reaction rates while non-extensive distributions have been used for $q=0.5$ (dotted line) and for $q=2$ (dashed line).  We see that in both cases, the $^7$Li abundance increases. This also occurs for values either below or slightly above $q=1$.  In Table \ref{tab1}, we show the BBN predictions for element abundance using  Maxwellian and non-Maxwellian distributions. All figures have the  same power of ten as in the observational data presented in the last column.

\begin{table}[htbp]
\vspace{0.0cm}
\centering
\caption{\label{tab:table1} BBN predictions for element abundance using  Maxwellian and non-Maxwellian distributions. All figures have the  same power of ten as in the observational data \cite{Ko11} presented in the last column \cite{BFH13}.}
\begin{tabular}{ccccccc}
\hline
\hline
 &Maxwell&Non-Max.&Non-Max.&Observation \\ 
 &   BBN  &   $q=0.5$             &$q=2$&\\ \hline

${^4}$He/H&0.249&0.243&0.141&$0.2561 \pm 0.0108$ \\
D/H&2.62&3.31&570&$2.82^{+0.20}_{-0.19}$($\times 10^{-5}$) \\
${^3}$He/H&0.98&0.91&69.1&$(1.1\pm 0.2)$($\times 10^{-5}$) \\
${^7}$Li/H&4.39&6.89&356.&$(1.58\pm 0.31) (\times 10^{-10})$ \\ \hline
\hline
\end{tabular} 
\vspace{0.0cm}
\label{tab1}
\end{table}     

We conclude that it is not possible to solve the lithium puzzle with use of a non-extensive statistics to calculate the reaction rates during the BBN. The departure from the BG statistics in fact worsens the lithium problem by increasing its abundance. The details for the reasons of this increase are explained in Ref. \cite{BFH13}. As seen in Table  \ref{tab1}, the abundances of the other elements are also substantially affected by a departure from the extensive statistics. A chi-square fit to the observations has revealed a very narrow window for the non-extensive parameter of the order of $q=1.00^{+0.05}_{-0.02}$ \cite{BFH13}.

\section{NUCLEAR CROSS SECTIONS UNCERTAINTIES}

The experimental values of the nuclear reaction cross sections of interest for nuclear astrophysics can be fitted using the  R-matrix method, where the resonances and background are obtained  by solving the Schr\"odinger equation  with matching conditions at a channel radius, or radii. The method leads to the reproduction of phase shifts and cross sections described in terms of a small set of parameters, and by extrapolation one obtains the cross sections at the low astrophysical energies. The matching condition at the channel radii leads to S-matrix poles at energies $E_\lambda$, and reduced widths, $\gamma_\lambda$. The energy dependence of the R-matrix becomes 
\begin{equation}
R_{ij}(E)=\sum_{\lambda =1}^N {\gamma_{\lambda i} \gamma_{\lambda j} \over E_\lambda - E},
\end{equation}
where $i$ and $j$ denote the reaction channels, with  momentum, $J$, and parity, $\Pi$. The reduced widths, $\gamma_{\lambda i}$  are related to solutions of the Schr\"odinger equation for each channel radius \cite{DB10}. In Ref. \cite{Piz14} R-matrix fits have been done to a collection of data based on direct and indirect measurements such as those obtained recently with the Trojan Horse Method (THM). 
Various nuclear reaction rates are among the most important input parameters for the BBN nucleosynthesis, but only the main 12  reactions listed in Table \ref{rent} have been considered \cite{KT90}. 

\begin{table}[htbp]
\vspace{0.0cm}
\begin{tabular}{|l|l|l|l|l|}
\hline
\hline
${\rm n} \leftrightarrow {\rm p}$  &${\rm p(n,}\gamma{\rm )d}$  & ${\rm d(p,}\gamma)^3{\rm He}$ & ${\rm d(d, p)t}	$ \\ \hline\hline
${\rm  d(d, n)}^3{\rm He}	$ &$ ^3{\rm He(n,p)t}$  &${\rm  t(d,n)}^4{\rm He}$ & $^3{\rm He(d, p)}^4{\rm He}  $  \\ \hline
$^3{\rm He}(\alpha,\gamma)^7{\rm Be}$  & ${\rm t(}\alpha,\gamma)^7{\rm Li}$ & $^7{\rm Be(n,p)}^7{\rm Li}$  &$^7{\rm Li(p,} \alpha )^4{\rm He}	$ \\ \hline
\hline
\end{tabular} 
\caption{The 12 most relevant nuclear reactions during the BBN. }\label{rent} 
\end{table}  

\begin{center}
\begin{figure}[t]
\includegraphics[width=8.5cm]{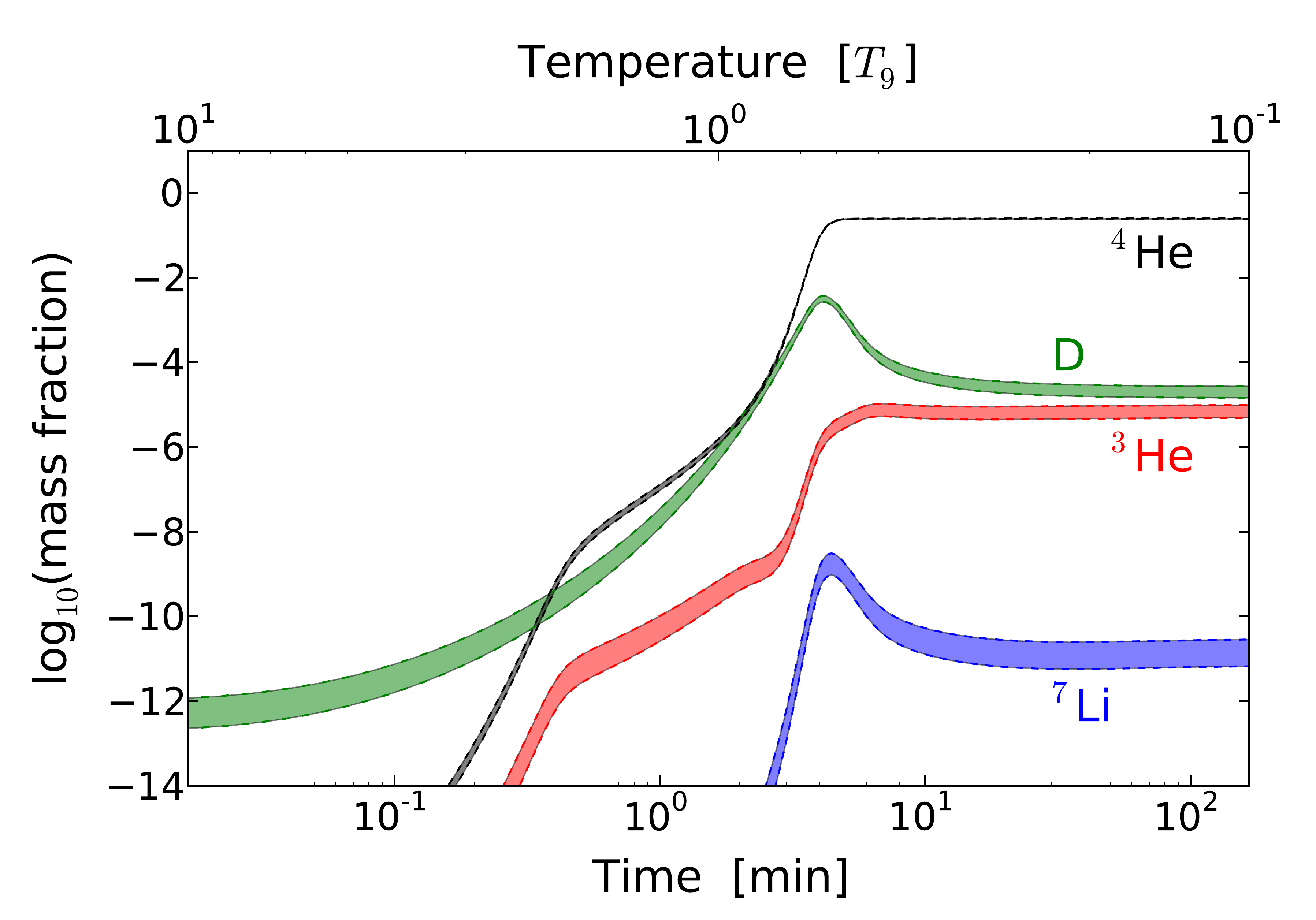}
\includegraphics[width=7.8cm]{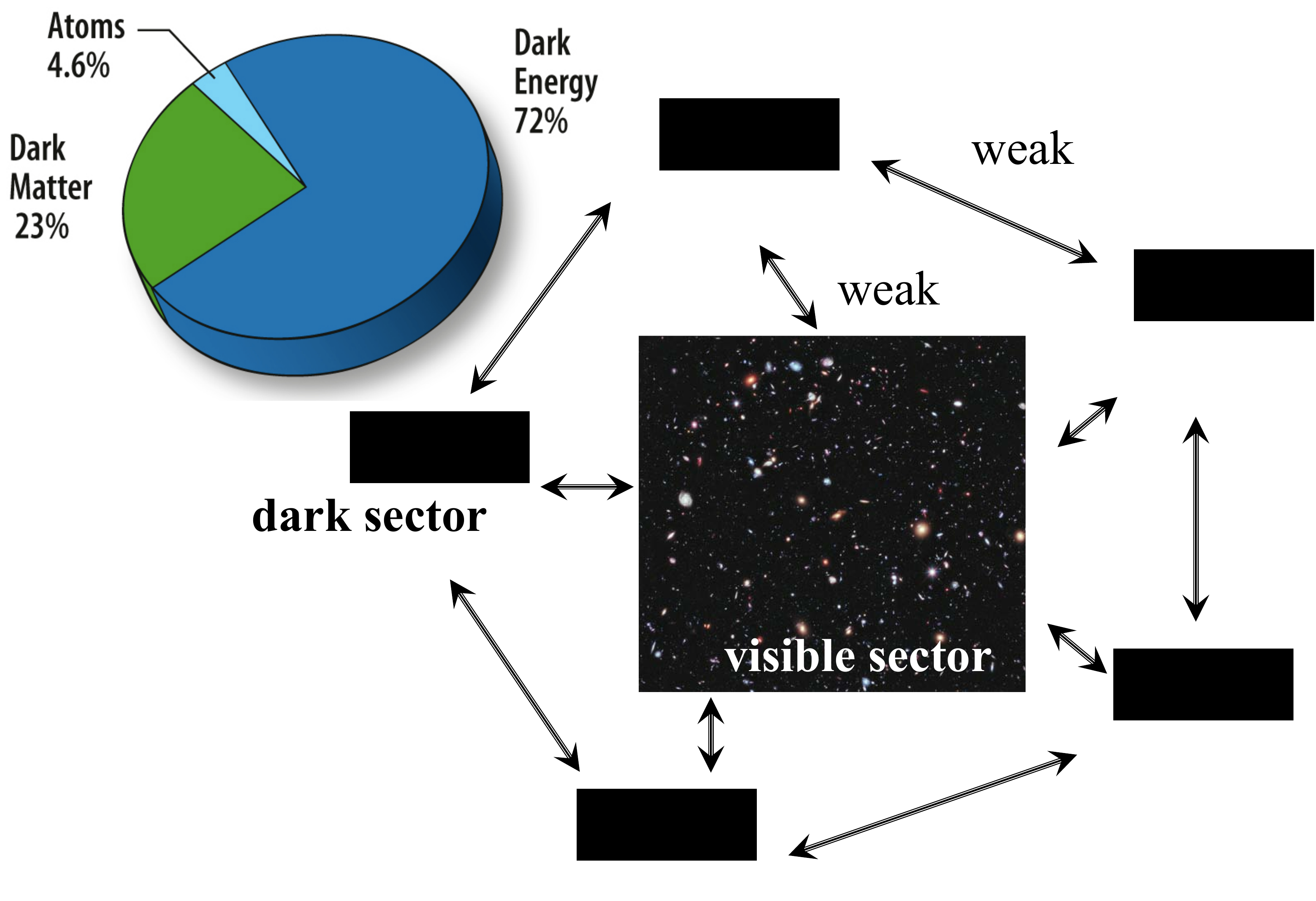}
\caption{{\it Left:} Calculated BBN abundances of $^3$He, $^4$He, D and $^7$Li as a function of time and temperature using best fits to existing experimental data \cite{Piz14}. The black curve is for the $^4$He mass fraction, the green curve is the deuterium abundance, the red curve is for the $^3$He abundance and blue curve for the $^7$Li abundance. Errors in the experimental data are manifest in the error bands around the abundance curves. (Adopted from Ref. \cite{Piz14}). {\it Right:} Visible matter accounts for only 5\% of the energy content in the Universe, while dark matter accounts for about 25\% of it. }\label{fig2}
\end{figure}
\end{center}

In Figure \ref{fig2} we show the calculated BBN abundance of $^3$He, $^4$He, D and $^7$Li as a function of time and temperature using best fits to existing experimental data \cite{Piz14}. The black curve is for the $^4$He mass fraction, the green curve is the deuterium abundance, the red curve is for the $^3$He abundance and blue curve for the $^7$Li abundance. Errors in the experimental data are manifest in the error bands around the abundance curves.  In Table \ref{tabbbn} the BBN calculations using fits to recent experimental  data for BBN reactions are compared with observations. Data for (a) are mass fraction for $^4$He from Ref. \cite{YT10}, (b)  deuterium abundance from the mean average $\left< ({\rm D/H})\right > = (2.82 \pm 0.26) \times 10^{-5}$, compatible with $\Omega_b h^2 \ ({\rm BBN}) = 0.0213 \pm 0.0013$ \cite{Mea06},  (c)  $^3$He abundances  from Ref. \cite{BRB02},  (d)  lithium abundance  from Ref. \cite{Sbo10}. It is clear from Figure \ref{fig2} and Table \ref{tabbbn} that the observed lithium depletion puzzle is not so evident when experimental errors in the measured S-factors, or cross sections, are included.

 More recently the second lithium problem, i.e., the disagreement between observations and BBN predictions for the lithium $^6$Li/$^7$Li isotopic ratio was discussed in Ref. \cite{MSB16}. While BBN predictions are $^6{\rm Li}/^7{\rm Li} \sim 10^{-5}$,  observations yield $^6{\rm Li}/^7{\rm Li} \sim 2 \times 10^{-2}$ \cite{Asp06}. In Ref. \cite{MSB16} a re-analysis of the reaction $^4$He$(\alpha,\gamma)^6$Li was performed, including new predictions for angular distribution. A nice agreement with recent experimental data from the LUNA collaboration \cite{LUNA} was found.  This result reinforces the BBN predictions for the lithium isotopic ratio and leads to a new value of $^6{\rm Li}/^7{\rm Li} = (1.5 \pm 0.3) \times 10^{-5}$ (see also Ref. \cite{LMA13}), which now seems to agree much better with the BBN predictions. We conclude that it is not at all impossible that the problems associated with  lithium abundances might not be due to astration of elements, but to accurate measurements combined with better theories for nuclear reactions.

\begin{table}[htbp]
\vspace{0.0cm}
\centering
\caption{\label{tab:table2} BBN calculations using fits to recent experimental  data for BBN reactions compared with observations. Data for (a) are mass fraction for $^4$He from Ref. \cite{YT10}, (b)  deuterium abundance from the mean average $\left< ({\rm D/H})\right > = (2.82 \pm 0.26) \times 10^{-5}$, compatible with $\Omega_b h^2 \ ({\rm BBN}) = 0.0213 \pm 0.0013$ \cite{Mea06},  (c)  $^3$He abundances  from Ref. \cite{BRB02},  (d)  lithium abundance  from Ref. \cite{Sbo10}.
}
\begin{tabular}{|c|c|c|c|c|c|c|c|}
\hline
\hline
Yields &   Calculation   & Observation\\  \hline
 
$Y_p$ &0.2485$^{+0.001}_{-0.002}$&$0.2565 \pm 0.006^{(a)}$ \\ \hline
D/H ($\times 10^{-5}$)&2.692$^{+0.177}_{-0.070}$&$2.82\pm 0.26^{(b)}$ \\ \hline
${^3}$He/H ($\times 10^{-6}$) &9.441$^{+0.511}_{-0.466}$&$\geq 11.\pm 2.^{(c)}$\\ \hline
${^7}$Li/H ($\times 10^{-10}$)&4.683$^{+0.335}_{-0.292}$&$1.58\pm 0.31^{(d)} $ \\ \hline
\hline
\end{tabular} 
\vspace{0.0cm}
\label{tabbbn}
\end{table}     

\section{PARALLEL UNIVERSES OF DARK MATTER}

Dark Matter (DM) comprises most of the matter in the universe and almost nothing is known about it, except for its gravitational properties
 \cite{Feng2010,bertone05,bertone10}. Except for gravity, it interacts very weakly with known baryonic matter, or perhaps not at all. Its existence is based on observations of galaxy clusters dynamics and anisotropies of the Cosmic Microwave Background (CMB). It is possibly composed of   hypothetic particles such as Weakly Interacting Massive Particles (WIMPs), supersymmetric particles, sterile neutrinos, etc.
 DM could be built up of mirror(s) sector(s) of particles \cite{Yang,Kobzarev,Pavsic,Foot,Akhmedov} with each mirror sector being composed of particle copies of the Standard Model (SM), but not exact copies with different masses and couplings. 

In Refs. \cite{Oli11,BFF12}  a mirror model of parallel universes was developed based on a not fully explored SU(3) gauge  symmetry for multiple universes. Based on the observation for the density parameters  $\Omega_{DM}$ and $\Omega_{b}$, there are about 5 times more dark matter than visible matter, more precisely, $\Omega_{DM} / \Omega_{b} =4.94 \pm 0.66$. One can assume that there are 5 dark sectors and one visible sector (see Figure \ref{fig2}, right). The dark sectors are exact replicas of the Standard Model (SM) of elementary particle physics.  In Ref. \cite{Oli11} one has introduced a new Weakly Interacting Massive Gauge Boson (WIMG) responsible for the coupling between all sectors, also with ordinary matter. The WIMG has to be a massive boson, leaving properties of the SM and gravity unchanged. It also has to be compatible with BBN results and CMB observations.  For energies much lower than the electroweak scale, particles are basically massless and one can group matter fields in terms of their intrinsic electric charges
\begin{equation}
Q_1 = \left( \begin{array}{c} u \\ c \\ t \end{array} \right) , \quad  Q_2 = \left( \begin{array}{c} d \\ s \\ b \end{array} \right) , \quad
Q_3 = \left( \begin{array}{c} e \\ \mu \\ \tau \end{array} \right) , \quad
 Q_4 = \left( \begin{array}{c} \nu_e \\ \nu_\mu \\ \nu_\tau \end{array} \right) \, ,
 \label{matter}
\end{equation}
where the notation $\mathcal{Q} = \left\{ Q_1, Q_2, Q_3, Q_4 \right\}$ will be used next.

A similar structure wcan be assumed for DM, with each DM sector $\mathcal{Q}_s$ having 4 multiplets and its own copy of the SM. Their 
electroweak sectors bosons will couple only within their own sectors. The WIMG gauge field $M^a_\mu$ couples to  matter fields $Q_{i,s}$, where $s$ denotes the  $N_Q=6$ sectors and $i$ denotes fermions, as in Eq. (\ref{matter}). The mass, $M^a_\mu$, of the WIMG is generated by a   real scalar field $\phi^a$, in an adjoint representation of the $SU(3)_Q$ group, with the condition that the WIMG  interaction is short ranged. The proposed Lagrangian in this gauge theory is
\begin{eqnarray}
 \mathcal{L}   =   - \frac{1}{4} F^a_{\mu\nu} F^{a \, \mu\nu}  + 
 \sum^{N_Q}_{s=1}\sum^4_{i=1} \overline Q_{i,s} \, \left\{i \gamma^\mu D_\mu -m_s\right\}\, Q_{i,s}
 + ~ \frac{1}{2} \left( D^\mu \phi^a \right) \left( D_\mu \phi^a \right) - V_{oct}( \phi^a \phi^a ) +  \mathcal{L}_{GF} +  \mathcal{L}_{gh},
 \label{lagrangeano}
\end{eqnarray}
where $ \mathcal{L}_{GF}$ fixes the gauge in the Lagrangian and $\mathcal{L}_{gh}$ carries the ghost terms, $D_\mu = \partial_\mu + i g_M T^a M^a_\mu$  is the covariant derivative, $T^a$ are the generators of $SU(3)_Q$ and $V_{oct}$ is the potential energy related to $\phi^a$. 
Note that the second term in Eq. (\ref{lagrangeano}) includes a sum over all families of fermions.
The Lagrangian  $\mathcal{L}$ is a sum of the SM in each sector with $s = 1, \cdots, N_Q$. Terms describing the  quantization of the theory are not shown.  The WIMG mass arises from the operator
$ (g^2_M/2) \, \phi^c (T^a T^b)_{cd} \phi^d M^a_\mu M^{b \, \mu} $, by assuming a non-vanishing boson condensate
$\langle \phi ^a \phi^b \rangle$ due to  local fluctuations of the scalar field.  With the additional assumptions that
$\langle \phi ^a \rangle = 0$ and $\langle \phi^a \phi^b \rangle = v^2 \delta^{ab}$, and using $\mbox{tr} (  T^a T^b )= 3 \, \delta^{ab}$, one gets for the WIMG mass the value $M^2  = 3 \, g^2_M   v^2 $.  The new  degrees of freedom will increase the early universe expansion rate  \cite{Berezhiani1996} with implications for the BBN. 

The radiation density during the BBN includes a number of new particles with are constrained by the $^4$He  abundance and the baryon-to-photon ratio $\eta_b$,  so that densities and entropies are given by
\begin{equation}
 \rho(T)=\frac{\pi^{2}}{30} \, g_*(T) \,T^4 \quad\mbox{and}\quad
 s(T)=\frac{2\pi^{2}}{45} \, g_s(T)\, T^3 ,
 \label{energy_entropy}
\end{equation}
where
\begin{equation}
g_*(T)=\sum_B g_B \left(\frac{T_{B}}{T}\right)^4 + \frac{7}{8} \sum_F g_F \left(\frac{T_{F}}{T}\right)^4, \ \ \ \ 
{\rm and} \ \ \ \
g_s(T)=\sum_B g_B \left(\frac{T_{B}}{T}\right)^3 + \frac{7}{8} \sum_F g_F \left(\frac{T_{F}}{T}\right)^3.
\end{equation}
These are the number of degrees of freedom during the BBN, $g_{B(F)}$ for  bosons (fermions),  $B(F)$, with temperatures $T_{B(F)}$.
$T$ denotes the temperature of the photon thermal bath.

If the dark sectors were decoupled, two temperatures would emerge: $T$ for ordinary matter and $T'$ for the dark sectors. The
energy $\rho'(T')$ and entropy $s'(T')$ densities for the dark sectors are given by Eqs. (\ref{energy_entropy}) with the substitutions $g_*(T) \rightarrow g'_*(T')$, $g_s(T) \rightarrow g'_s(T')$, and  $T\rightarrow T'$. Separate conservation of entropy in visible and dark sectors yields a time independent parameter $x=(s'/s)^{1/3}$. If each dark sector has the same matter content as in the visible sector,   $g_s(T_0) = g_s^\prime(T'_0)$, leading to $x=T'/T$. The Friedman equation for a radiation dominated epoch is  $ H(t)=\sqrt{\left(8\pi/3 c^2\right) \, G_{N} \, \bar{\rho}}$, with the total energy density  $\bar{\rho} = \rho\, + \, N_{DM} \, \rho'$, where
$N_{DM} = N_Q - 1$ is the number of dark sectors. Using the relation (\ref{energy_entropy}) for the density in the dark sectors, $\rho'$, leads to
\begin{equation}
H(t)=1.66 \, \sqrt{\bar{g}_{*}(T)} \, \frac{T^2}{M_{Pl}},
\ \ \ \ \ \
{\rm where}
\ \ \ \ \
\bar{g}_{*}(T) = g_{*} (T) \left( 1+ N_{DM} \, a \, x^4 \right),
\end{equation}
where $M_{Pl}$ is the  Planck mass and $a = \left(g'_{*}/g_{*}\right) \left(g_{s}/g'_{s}\right)^{4/3}\sim 1$, for a not too small $T'/T$  \cite{Berezhiani1996}. At a BBN  temperature of about 1 MeV, the degrees of  freedom of photons, electrons, positrons and neutrinos are in quasi-equilibrium and $g_{*}(T=1 \ {\rm MeV}) = 10.75$.
Due to the additional dark particles,  $g_{*}$ changes to $\bar{g}_{*} = g_{*} \left(1 + N_{DM} \, x^4 \right)$. If one tries to explain the
$^{4}$He, $^{3}$He, and D abundances, one gets \cite{Oli11} $T' / T < 0.78 / N^{1/4}_{DM}$. With $N_{DM} = 5$, this yields $T' /T < 0.52$, i.e.,  dark universes have to be always  colder than our universe if the observed element abundances are to be reproduced.
Other interesting features of the model is that the baryon-to-photon ratio for the dark sectors comes at as  $\eta'_b \sim 2.1 \, N^{3/4}_{DM}\eta_b$. Using  $N_{DM} = 5$ one finds $\eta'_b\sim 7\eta_b$, i.e., BBN runs very differently in the visible and in the dark and cold sectors. The additional degrees of freedom were forced to reproduce the observed abundances in the visible universe. Ref. \cite{Oli11} proved that such a model is compatible with what is observed in the visible universe. But turning the argument around and assuming that the similar conditions in the dark sectors were varied, the model would predict different temperatures, baryon-to-photon ratios, and other changes in the visible universe. The impact of these changes on BBN in our visible universe are entirely open to imagination.  The lithium problem might find a solution in this way, but at this point in history we cannot really ascertain what would be the physical properties of the dark sectors and what would they mean if no other probes than gravity and a highly hypothetical WIMG would exist.

\section{CONCLUSIONS}

In this work we have discussed a few studies by our group at Texas A\&M University (Commerce \& College Station) on the cosmological lithium problem. The goal is by no means to make a comprehensive review of this research field.  For this purpose, there exist much more authoritative reviews in the literature (see, e.g., Refs. \cite{Stei07,Fie11}). 

The lithium puzzle is associated with the BBN prediction of $^7$Li being larger than what is observed. The puzzle recalls an equivalent situation with the solar neutrino problem many decades ago. It might be that there is no puzzle at all. The solution might lie in the way that lithium is processed in stars. The solution might as well be in the realm of nuclear physics. As we have discussed in this text, another  lithium problem, i.e., the large difference between BBN predictions and observations for the isotopic ratio $^6$Li/$^7$Li, might find its solution in the more precise experiments combined with better nuclear reaction theory \cite{MSB16}.   Similar cases such as the electron screening puzzle mentioned earlier have recently found a likely solution also in the realm of nuclear physics. For example,  as a nuclear clustering phenomena in light nuclei, instead of a solution in atomic physics.  Since $^7$Li is produced mainly via the electron capture process  $^7{\rm Be} + e^- \rightarrow ^7{\rm Li} + \nu_e$, the destruction  of $^7$Be via several channels could also become a possible solution for the puzzle \cite{Brog12}. The fact is that to the present date there is  no solid key for this puzzle that has survived detailed tests in theory and experiment. As in the case of the solar neutrino problem, with Pontecorvo's proposition of neutrino oscillations \cite{Pon57},  the answer for the cosmological lithium problem might already have been found and is waiting out there for a confirmation. 

\section{ACKNOWLEDGMENTS}
C.A.B. Acknowledges support from by the U.S. NSF Grant No. 1415656 and the U.S. DOE Grant No. DE-FG02-08ER41533. A. M. M.  acknowledges that this material is based upon work supported by the U.S. Department of Energy,  Award Number DE-FG02-93ER40773, by the U.S. Department of Energy, National Nuclear Security Administration,  Award Number DE-FG52-09NA29467 and  by the US National Science Foundation under Award PHY-1415656. O.O. acknowledges financial support from FCT under contract PTDC/FIS/100968/2008.


\end{document}